\documentclass[11pt,a4paper]{article}

\usepackage{jheppubmod}

\usepackage{graphicx}
\usepackage{amssymb}
\usepackage[mathscr]{eucal}
\usepackage{amsmath}
\usepackage{afterpage}
\usepackage{slashed}
\usepackage{feynmp}
 \usepackage[table]{xcolor}
 \usepackage{url}
 \usepackage{hyperref}
\hypersetup{colorlinks=true,linkcolor=magenta,anchorcolor=green,citecolor=cyan,filecolor=black,menucolor=black,urlcolor=brown}

\newcommand{\ms}{m_\star}

\newcommand{\Mp}{M_{_\mathrm{Pl}}}

\newcommand{\gsimm}{\raise.3ex\hbox{$>$\kern-.75em\lower1ex\hbox{$\sim$}}}
\newcommand{\lsimm}{\raise.3ex\hbox{$<$\kern-.75em\lower1ex\hbox{$\sim$}}}

\newcount\hh
\newcount\mm
\mm=\time
\hh=\time
\divide\hh by 60
\divide\mm by 60
\multiply\mm by 60
\mm=-\mm
\advance\mm by \time
\def\thistime{\number\hh:\ifnum\mm<10{}0\fi\number\mm}

\newcommand{\be}{\begin{equation}}
\newcommand{\ee}{\end{equation}}
\newcommand{\ba}{\begin{eqnarray}}
\newcommand{\ea}{\end{eqnarray}}

\newcommand{\bea}{\begin{eqnarray*}}
\newcommand{\eea}{\end{eqnarray*}}

\title{$R^2$ Dark Energy in the Laboratory}

\author[a]{Philippe Brax,}
\author[a]{Patrick Valageas,}
\author[a,b]{\textrm{and} Pierre Vanhove}
\affiliation[a]{Institut de Physique Th\'eorique, Universit\'e Paris-Saclay, CEA,CNRS,\\
F-91191Gif sur Yvette, France }
\affiliation[b]{National Research University Higher School of Economics, Russian Federation}
	
\abstract{We analyse the role, on large cosmological scales and
  laboratory experiments, of the leading curvature squared contributions to the
  low energy effective action of gravity.  We argue for a natural
  relationship $c_0\lambda^2\simeq 1$ at low-energy between the ${\cal R}^2$
  coefficients $c_0$ of the Ricci scalar squared term in this expansion and the  dark energy scale
  $\Lambda=(\lambda\,\Mp)^4$ in four dimensional Planck mass units.
  We show how the compatibility between the acceleration of the
  expansion rate of the Universe, local tests of gravity and the
  quantum stability of the model all converge to select such a
  relationship up to a coefficient which should be determined
  experimentally. When embedding this low energy theory of gravity
  into candidates for its ultraviolet completion, we find that the
  proposed relationship is guaranteed in string-inspired supergravity models with modulus
  stabilisation and supersymmetry breaking leading to de Sitter compactifications. In this case,
  the  scalar degree of freedom of ${\cal R}^2$ gravity is associated to a volume
  modulus. Once written in terms of a scalar-tensor theory, the effective theory
  corresponds to a massive scalar field coupled with the universal strength
  $\beta=1/\sqrt{6}$ to the matter stress-energy tensor.  When the
  relationship $c_0\lambda^2\simeq 1$ is realised we find that on
  astrophysical scales and in cosmology the scalar field is ultralocal
  and therefore no effect arises on such large scales. On the other
  hand, the scalar field mass is tightly constrained by the
  non-observation of fifth forces in torsion pendulum experiments such
  as E\"ot-Wash. It turns out that the observation of the dark energy
  scale in cosmology implies that the scalar field could be detectable
  by fifth force experiments in the near future.

}

\begin{document}
\preprint{IPhT-t17/066}
\maketitle
\flushbottom

\section{Introduction}

The acceleration of the expansion of the Universe~\cite{Copeland:2006wr,Clifton:2011jh} has not received any
natural explanation within Quantum Field Theory yet. The modern way of
describing the problem of the acceleration of the Universe is
two-pronged. First there is the ``old"' cosmological constant problem~\cite{Weinberg:1988cp}
whereby all the massive particles of the Universe contribute to the
vacuum energy with threshold corrections which are quartic in their
masses. Phase transitions, and at least the electro-weak and the
quantum chromodynamics ones, also contribute to an alarming
level. These large contributions must cancel in order to match with
observations. This could for instance arise from some hidden
underlying symmetry which could relate in some unknown way the
spectrum of observed and unobserved particles.  One popular mechanism
relies on supersymmetry, but some cancellations are known as well to
occur to lowest order in perturbation theory in non supersymmetric
models~\cite{Kachru:1998hd,Kachru:1998pg,Harvey:1998rc,Blumenhagen:1998uf,Angelantonj:1999gm}.
In these models the misaligned bosonic and fermionic spectrum still
cancel in radiative corrections. At low energy and assuming that such
a solution to the ``old'' cosmological constant problem can be
postulated, one must face the ``dark energy'' problem whereby the
measured vacuum energy obtained from cosmological observation
$\rho_{\Lambda 0}= \lambda^4 \Mp^4 = 3\Omega_{\Lambda} \Mp^2 H_0^2$ is tiny and of
order~\cite{Olive:2016xmw}
\begin{equation}\label{e:lambdadef}
\lambda = (3 \Omega_{\Lambda 0})^{1/4} ( H_0/\Mp )^{1/2}
\simeq (3\times 10^{-122}
)^{1\over4}\simeq 4\times 10^{-31} .
\end{equation}
We have introduced $\Omega_{\Lambda 0}\sim 0.7$ as the fraction of dark energy in the Universe and $\rho_c= 3\Mp^2 H_0^2$ is the critical density of the Universe.

In this paper, we do not introduce a new field which would mimic the late time behaviour of a Universe as dominated by a constant vacuum energy. On the contrary we take known physics seriously and take into account the gravitational
corrections to the Einstein-Hilbert action which would arise from high energy physics. In essence, we assume that the ``dark energy'' problem has its solution within the gravitational sector of the theory describing low energy physics.
We consider the (parity-even) low-energy effective action depending on the curvature tensor (our metric
has signature $(-+\cdots+)$)
\begin{equation}\label{EH-R2}
\mathcal S= \int d^4x \sqrt{-g} \, \left(-\lambda^4
  \Mp^4+{\Mp^2\over2}\mathcal R+ c_0\,\mathcal R^2-c_2\,
  \left(R_{\mu\nu}R^{\mu\nu}-\frac13 \mathcal R^2\right)+\cdots \right)  \, ,
\end{equation}
and we argue in favour of  a natural relation between the small
cosmological constant $\lambda$ and the $\mathcal R^2$
coefficient
\begin{equation}\label{e:MainRel}
    c_0\simeq \lambda^{-2} \simeq 10^{61}\,,
\end{equation}
in the gravitational sector of the low-energy effective action.

In section~\ref{sec:R2} we discuss the general setup for this work in
the context of curvature squared  models. We explain  the important
difference between the $\mathcal R^2$ and $R_{\mu\nu}^2$ terms which
require that their respective coefficient satisfy the hierarchy $c_0\gg
c_2$. In section~\ref{sec:observations}
we derive the upper bound $c_0\lesssim \lambda^{-2}$ from
observational constraints by using the equivalence between the scalar
curvature squared models and a massive scalar coupled
with the universal strength $\beta=1/\sqrt{6}$ to the matter
stress-energy tensor~\footnote{Such a low value of the coupling corresponds to a suppression scale for the operators involving both matter and the scalar $M_{\rm sup}= M_{\rm Pl}/\beta$ which is larger than the Planck scale, implying that
the particle physics bound on $M_{\rm sup}\gtrsim 1$ TeV is easily evaded \cite{Brax:2009aw}.}. In section~\ref{sec:Theoretical-constraints} we
show that quantum stability of the previous model implies the lower
bound $ \lambda^{-2}\lesssim c_0$.  In section~\ref{sec:EFT} we
provide various  effective field theory constructions where the
relationship~\eqref{e:MainRel} is discussed  in string theory
compactifications with no warping in section~\ref{sec:string} and in string-inspired
supergravity models in
section~\ref{sec:sugra}. In this  section~\ref{sec:nilpotent} we derive the
relation~\eqref{e:MainRel} by considering a model  with warping and brane-antibrane supersymmetry breaking. Section~\ref{sec:conclusion} contains our
conclusion. The appendix~\ref{sec:labo} is devoted to a review of
laboratory tests of modified gravity, appendix~\ref{sec:astro}
to a review of the astrophysical tests and appendix \ref{sec:kah} to supergravity details.

\section{${\cal R}^2$-type models and their status}
\label{sec:R2}

In this paper, we investigate extensions of General Relativity where
the Einstein-Hilbert action is supplemented by the curvature squared
terms in~\eqref{EH-R2}.  This action should be understood as a low
energy effective action, which arises from a more fundamental
ultraviolet complete theory. We are not taking into account  the
Gauss-Bonnet term which does not contribute as it is a
total derivative, unless it  is field dependent. We assume that its
coefficient is constant at the low energies considered in this
 work and therefore the Gauss-Bonnet term can be safely discarded.
The omitted terms in the curvature expansion must be suppressed
 compared to the leading curvature squared terms. In string models, we argue that
 this is the case and that they are suppressed by the string scale and
 become irrelevant at low energy in the small curvature regime which
 describes the dynamics of the Universe.\footnote{The largest
   gravitational potential on cosmological scales for the largest
   galaxy clusters is no more than $10^{-4}$.}
In section~\ref{sec:EFT} we will consider
supersymmetry breaking compactifications of string theory. Therefore,
the parameters $\lambda$, $M_{\rm Pl}$, $c_0$ and $c_2$
are expected to be correlated, as they should be expressed in terms of
the basic parameters of the more fundamental theory.  In particular,
we might expect the parameters $c_0$ and $c_2$ to be of the same order
although this leads to issues which will be detailed below.  The scale
$M_{\rm Pl}$ is constrained to be the usual Planck mass, in order to
be consistent with the measurements of Newton's constant.  We assume
that the first term, which has been identified with the observed
vacuum energy, has the same origin as the other terms.  All these
requirements can be understood as a constraint on the parameters of
the ultraviolet complete theory by requiring the action~\eqref{EH-R2}
to arise in the infrared.  We shall give examples in subsequent
sections of the derivation of all these terms from more fundamental
actions, e.g. from string theory and $\mathcal N=1$ supergravity.

The tiny value~\eqref{e:lambdadef} of the dimensionless coefficient
$\lambda$ is presumed to arise from a wide hierarchy of scales in the
more fundamental theory that leads to the low-energy effective action
(\ref{EH-R2}). At low energy and considering the one-loop contribution of massive particles, the vacuum energy can be written as
\be
\lambda^4 M_{\rm Pl}^4\simeq 2\Lambda_0 M_{\rm Pl}^2 + \rho_{\rm trans} + \sum_{j} (-1)^{2j} (2j+1) \frac{m_j^4}{64\pi^2 }\ln \frac{m_j^2}{\mu^2}
\label{cos}
\ee
where the first term is the bare cosmological constant, $\rho_{\rm trans}$ corresponds to the energy density associated to all the phase transitions such
as QCD confinement or the electroweak phase transition and finally the last term summed over all particles of spin $j$ is the quantum contribution from vacuum fluctuations. Typically
$\lambda$ is renormalised at a scale $\mu_0$ and identified with the measured value obtained from cosmological observations. This implies that at a different scale $\mu$ we have
\be
\lambda^4 (\mu)= \lambda^4 (\mu_0)+ \sum_{j} (-1)^{2j} (2j+1) \frac{m_j^4}{64\pi^2 M^4_{\rm Pl} }\ln \frac{\mu_0^2}{\mu^2}.
\ee
In cosmology the meaning of $\mu$ is not as clear as in particle physics where it can be identified with the centre of mass energy for instance. Here $\mu$ may be identified with
the Hubble rate or any other cosmological scale related to the evolution of the Universe. If this is the case then low energy stability of the vacuum energy requires that
\be
\sum_{j} (-1)^{2j} (2j+1) \frac{m_j^4}{64\pi^2 M^4_{\rm Pl}} \simeq 0
\ee
at a level of accuracy as good as the tiny value of $\lambda (\mu_0) =: \lambda$. This is the ``old" cosmological constant problem.

In the following, we will explicitly follow a different route and separate the description of the vacuum energy into two  blocks. The first one concerns matter. We assume that in (\ref{cos})
all the matter dependent contributions vanish or contribute in a negligible way to the vacuum energy at the energy $\mu$ associated to our present Universe.
This leaves then the gravitational sector of the theory free and responsible for the generation of ``dark energy". To do this, we separate a piece of the  bare cosmological constant which will be identified
with the classical energy density $V_{\rm vac}>0$ in the gravitational sector, for instance from a de Sitter string compactification independently of the matter content of the Universe
\be
 2\Lambda_0 M_{\rm Pl}^2= 2\tilde \Lambda_0 M_{\rm Pl}^2 +V_{\rm vac}
\ee
and  we require the following tuning
\be
2\tilde \Lambda_0 M_{\rm Pl}^2 +\rho_{\rm trans} + \sum_{j} (-1)^{2j} (2j+1)\theta(m_j-\mu_0) \frac{m_j^4}{64\pi^2 }\ln \frac{m_j^2}{\mu^2}=0
\label{tune}
\ee
where the scale $\mu_0$ is chosen such that all the massive matter particles contribute to the sum.  In a sense this corresponds to a tuning of the bare cosmological constant $\tilde \Lambda_0$ which acts as a counter term for the contributions of all the massive particles and the phase transitions of the early Universe.
This leaves aside   the loop contributions from the massive degrees of freedom in the gravitational sector at low energy, typically from massive scalars or gravitons which survive at scales  below $\mu_0$ above which
all the matter particles have been integrated out. All in all we have
\be
\lambda^4 M_{\rm Pl}^4= V_{\rm vac} + \sum_{j} (-1)^{2j} (2j+1) \theta ( \mu_0-m_j) \frac{m_j^4}{64\pi^2 }\ln \frac{m_j^2}{\mu^2}
\label{cos1}
\ee
where no matter particle is present in the sum, i.e. $\mu_0$ could be of the order of the electron mass,  apart from the neutrinos whose contributions may be of the right
order of magnitude if the neutrino masses were in the
milli-electronvolt (meV) range.

 The existence of the tiny coefficient $\lambda$
implies that the other coefficients $c_i$ in the gravitational sector of the theory are not necessarily of order
unity, as their value may also reflect this large hierarchy.  A
priori, they could be very small as well as very large, as the
combination that enters $c_i$ may be different from the one that
enters $\lambda$.  Parametrically, we write
\begin{equation}
c_i \propto \lambda^{-a_i} \qquad\textrm{with}\qquad i=0,2,
\label{ci-lambda}
\end{equation}
where the exponents $a_i$ would depend on the details of the fundamental
theory.
The coefficients $c_0$ and $c_2$
 are loosely constrained by observations,  the best bound arising from the
E\"ot-Wash experiments~\cite{Adelberger:2003zx}  (see as well~\cite{Donoghue:1994dn,Naf:2010zy,Berry:2011pb}) are $c_i\leq 10^{61}$.\footnote{We would like to
  thank John Donoghue for having informed us that a
  reanalysis of the upper bound leads to  a slightly lower bound than  the
  one given in~\cite{Donoghue:1994dn}.}
We will investigate the values of $c_i$, or of the exponent $a_i$, from the
points of view of both observations and theory, and we argue that  $a_0=2$,  i.e.
\begin{equation}\label{a0-2}
c_0 \propto \lambda^{-2}
\end{equation}
is favoured by the combination of experimental bounds and
stability under the Coleman-Weinberg
quantum corrections analysed in section~\ref{sec:Theoretical-constraints}.
Moreover one can realise this relation by choosing parameters
in  simple ultraviolet completions of the ${\cal R}^2$
models such as string compactifications or low energy
supergravity.

The model~\eqref{EH-R2} contains a scalaron of spin zero and mass
determined by $m_0^2= \Mp^2/(12\,c_0)$ and a massive spin-2 ghost
$m_2^2=\Mp^2/c_2$~\cite{Stelle:1976gc,Stelle:1977ry}. Clearly the
coefficients $c_0$ and $c_2$ have to be positive, otherwise the vacuum around which the curvature expansion is performed would not be a true vacuum.
 These masses are
parametrically of order $m_i\simeq \lambda^{a_i/2}\, \Mp$ up to a
numerical coefficient which is not calculable and depends on the
details of the physical origin of the models.
 The  two curvature squared terms in~\eqref{EH-R2} need to be treated
 differently as we now explain.

The Ricci scalar squared term $\mathcal R^2$ is equivalent to a massive scalar
 field, the scalaron. This scalaron is not a new field. It naturally appears in the
equivalence between  the gravitational $f(\mathcal R)$ theories and  scalar-tensor theories
with a coupling to matter which is universally
 fixed to $\beta=1/\sqrt6$ as we will discuss in
 section~\ref{sec:observations}.
As this scalaron couples to matter with a coupling strength
$\beta=1/\sqrt 6$, this would lead to the existence of a fifth
force of range $(\lambda^{a_0/2} \Mp)^{-1}\sim 82\ \mu {\rm m}$ for
$a_0=2$ which should be detectable in laboratory tests of short range
gravitational interactions such as
E\"ot-Wash~\cite{Hoyle:2004cw,Kapner:2006si}. On large cosmological
and astrophysical scales, the scalaron has virtually no effect as its
effective description is essentially ultra local~\cite{Brax:2016vpd}
on scales larger that $m_0=\sqrt{H_0\,\Mp}\simeq 10^{-3}$~eV.

 The main part of this work deals with the massive scalar
arising from the curvature squared term. But in an effective action
description the other curvature squared terms appear and one would
need to consider the massive spin-2 which is a ghost. The appearance
of this ghost is due to the truncation of the effective theory. As a result we must  consider plausible ghost-free ultraviolet completions of the curvature
squared terms as will be discussed in section~\ref{sec:EFT}.

For $a_2=2$ and $c_0\simeq c_2$ the massive spin-2 field has a mass of
the order $m_{2}\simeq 10^{-3}$~eV. The observational
bounds~\cite{Olive:2016xmw} stating that the mass of the graviton
should be less than $1.2\times10^{-22}~$eV would be evaded if the
massive spin-2 did not directly couple to matter
fields. Unfortunately, this is not the case
here~\cite{Stelle:1977ry}. On the other hand, the constraint on the
mass of the graviton is only applicable for gravitons of low energy
corresponding to frequencies of order $10-100$
Hz~\cite{Yunes:2016jcc,deRham:2016nuf}. The typical energy of such
gravitons is much smaller than $m_{2}$. Hence, the massive gravitons
that we consider here are of larger energy and would correspond to
excitations of frequencies much higher than the observation
window. Therefore the low energy theory of gravity with a massive
graviton of mass $m_{2}$ evades the gravitational wave
constraint.\footnote{We review the laboratory constraints in the
  appendix~\ref{sec:labo} and the astrophysical constraints
  in~\ref{sec:astro}.}  Similarly, one can safely work in cosmology
with the truncated Lagrangian at second order in the curvature as the
effective field theory of gravity at energies well below the MeV
scale. Indeed this is the range of energy scales where the presence of
a ghost can be tamed and kept below observational
bounds~\cite{Cline:2003gs}.  In terms of cosmology and Hubble rates,
this implies that the low energy expansion is safe from much earlier
than Big Bang Nucleosynthesis
($H_{\rm BBN} \lesssim 10^{-23} {\rm GeV}$). On the other hand
higher-order terms have to come into play before the inflationary
stage ($H_{\rm I} \sim 10^{13} {\rm GeV}$). Otherwise the expansion
becomes meaningless as involving a massive ghost in an energy range
where the vacuum becomes unstable.

As a result, in the event of $c_0\simeq c_2$, there would be a large gap
between the MeV scale where the presence of the ghost begins to be
felt and the large scale where the higher order terms cure the ghost
instability.\footnote{Unless the higher order terms become relevant at
  the MeV scale corresponding to new gravitational physics at very low
  energy. We discard this possibility which, in the case of string
  theory, would have implied a tower of Kaluza-Klein modes in the MeV
  range which has clearly not be seen at the LHC.}  In this energy
range the ghost instability would be lethal. A more natural
possibility would be that the only low energy degree of freedom in the
gravitational sector is the scalar field associated with a large value
of $c_0$ whilst the mass of the would-be spin two ghost would be
rejected towards scales of the order of the one where the higher order
operators become crucial. In this case, the spin 2 ghost would be an
artefact of the truncation with no physical consequences. This would
happen if
\begin{equation}\label{e:hierarchy}
c_0\gg c_2\,.
\end{equation}
We will argue that this could happen in the context of brane models of
string compactification where the massive spin 2 field would acquire a
fiducial mass of the order of the string scale and therefore would play no
role at low energy,  see section~\ref{sec:nilpotent}.
Consistency of the effective action requires that the mass $m_2$ of
the ghost is at least of the order the string mass scale. In this case the
ghost appears in regime where the perturbative expansion is not
reliable and becomes harmless (see for
instance~\cite[footnote~p54]{Gross:1986mw}).

It will turn out that the hierarchy~\eqref{e:hierarchy} between the curvature squared terms
in~\eqref{EH-R2} can be realised  thanks to  the equivalence between the Ricci scalar squared
term and the scalaron field.  In conclusion our analysis of the ``dark energy"  problem within the gravitational sector  requires
the hierarchy~\eqref{e:hierarchy} between the curvature square terms in the
effective action. It also necessitates the suppression of all the higher order derivative terms, something which can be
achieved in models where another fundamental scale such as the string scale, see~\ref{sec:string}, is present.


\section{Observational constraints}
\label{sec:observations}

We first consider the observational constraints on the coefficients of the action
(\ref{EH-R2}) that are associated with the additional scalar degree of freedom
entailed by this extension of General Relativity.
Thus, we focus on the Ricci scalar squared part
\begin{equation}
\mathcal S = \int d^4x \sqrt{-g}\,\left( -{\lambda^4\Mp^4} +\frac{\Mp^2}{2} \mathcal R
 + c_0 \mathcal R^2\right) .
\label{R2-action}
\end{equation}
This belongs to the class of $f(\mathcal R)$ models which have been studied in
great details in recent years and  where the Lagrangian is a  function of the scalar curvature, i.e.
$\mathcal L = {\Mp^2\over 2} f(\mathcal R)$, with
\begin{equation}
f(\mathcal R) =  -{2\lambda^4\Mp^2} +\mathcal R + {2c_0\over \Mp^2} \mathcal R^2 .
\label{fR-def}
\end{equation}
As is well known, this action can be transformed to the Einstein frame,
with the standard Einstein-Hilbert term supplemented by the Lagrangian
of a scalar field $\phi$ of the form
\begin{equation}
S_{\phi} = \int d^4x \sqrt{-g} \left[ - \frac{1}{2} (\partial\phi)^2 - V(\phi)
- \left( e^{\beta \phi/M_{\rm Pl}} - 1 \right) \rho \right] ,
\label{S-phi}
\end{equation}
with the parametric definition
\begin{align}
V(\phi) &= \frac{\Mp^2}{2} \frac{\mathcal R f'(\mathcal R) -
          f(\mathcal R)}{f'(\mathcal R)^2},\cr
e^{-2\beta\phi/\Mp} &= f'(\mathcal R), \qquad
\beta = \frac{1}{\sqrt{6}}\,.
\label{fR-phi}
\end{align}
This gives
\begin{equation}
V(\phi) =  \lambda^4 \Mp^4 e^{4\beta\phi/\Mp} + \frac{\Mp^4}{16 c_0}
\left( 1 - e^{2\beta\phi/\Mp} \right)^2 \, .
\label{e:Vscal}
\end{equation}
This normalization ensures the scalaron $\phi$ has a canonically normalised
kinetic term in the Einstein frame. Notice that it is only for the
special value of $\beta=1/\sqrt6$ that the theory is equivalent to
purely gravitational theory without any  extra scalar field introduced in the  model (see~\cite[\S
III.A]{Sotiriou:2008rp} for discussion of the relation of this
construction to Brans-Dicke models which can allow for  different values of $\beta$).
It also determines the universal coupling between the scalaron and the trace of the matter
stress energy potential, and in eq.~\eqref{S-phi} we used $T^{\mu}{}_{\mu}=-\rho$
for non-relativistic matter.
This coupling to matter gives rise to a fifth force, associated with the gradients of
the scalar field (i.e. its spatial inhomogeneities).

 We recover General Relativity in the effective action (\ref{R2-action}) in the low
curvature regime, when
\begin{equation}
c_0 \mathcal R\ll \Mp^2 , \;\;\; \mbox{hence} \;\;\; \phi\ll \Mp .
\end{equation}
For $c_0 \simeq 10^{61}$ this gives ${\mathcal R}^{1/2} \ll 10^{-12} {\rm GeV}=10^{-3}$~eV.
This is valid up to high redshift, much beyond the Big Bang Nucleosynthesis,
and for all astrophysical objects.
In this regime, we have that
\begin{equation}
    1+{4c_0\over \Mp^2} \mathcal
    R\simeq 1-2\beta{\phi\over \Mp}+{\mathcal O}(\phi^2/\Mp^2)
\end{equation}
hence
\begin{equation}
 \mathcal    R\simeq-{\beta\over2c_0}\, \phi\,\Mp +{\mathcal O}(\phi^2/\Mp^2)\,,
\end{equation}
and  the expansion to  leading order
of the potential
\begin{equation}
V(\phi) \simeq \lambda^4\Mp^4 +  4\lambda^4 \beta \Mp^3 \phi
+ \frac{\beta^2}{4} \left( 32
  \lambda^4 + \frac{1}{c_0} \right)
\, \Mp^2 \phi^2 +\mathcal O(\phi^3\Mp).
\end{equation}
This gives a massive scalar field Yukawa-coupled to matter.

In the presence of matter of density $\rho \sim m_\psi \bar \psi
\psi$ for non-relativistic fermions, the effective potential
$V_{\rm eff} = V + \delta V_{\textrm{matter}}$ for the scalar in the Einstein frame
is corrected by
\begin{equation}
\delta V_{\textrm{matter}}(\phi) = \rho \left( e^{\beta \phi/\Mp} -1 \right)
\simeq \rho \left( \frac{\beta \phi}{\Mp} + \frac{\beta^2 \phi^2}{2\Mp^2} \right) +\mathcal O(\phi^3/\Mp^3) .
\end{equation}
This is a chameleon model with a minimum $\phi_\star(\rho)$ for the scalaron
in the presence of matter and the effective mass $\ms (\rho)$,
\begin{equation}
\left. \frac{\partial V_{\rm eff}}{\partial\phi} \right |_{\phi_\star} = 0 \, , \;\;\;
\ms^2 = \left. \frac{\partial^2 V_{\rm eff}}{\partial\phi^2} \right |_{\phi_\star} \, .
\end{equation}
Since the matter density is much bigger that the
dark energy density $\rho \gg \rho_{\Lambda 0}=(\lambda \Mp)^4$ and
very small compared to the scalar curvature squared coefficient
$\rho/\Mp^4 \ll 1/c_0$ we obtain
\begin{equation}\label{e:vev}
\frac{ \phi_\star}{\Mp} \simeq -{2\over \beta}  {\rho\over \Mp^4}
c_0 + \mathcal O(\rho^2c_0^2/\Mp^8) \, , \;\;\;
\ms^2 \simeq \frac{\beta^2 \Mp^2}{2 c_0} +  \mathcal O(\rho/\Mp^2) .
\end{equation}
Because the coupling $\beta$ is of order unity, which would give deviations from
Newtonian gravity of order unity, the range $\ms^{-1}$ of the fifth force must be
below $0.1$ mm to satisfy the bounds from the E\"ot-Wash experiments,
as we recall in appendix~\ref{sec:Eot-Wash}.
This yields $c_0 \lesssim 10^{61}$, which satisfies $\rho c_0/\Mp^4 \ll 1$,
\begin{equation}
\phi_\star \simeq - \frac{2 c_0}{\beta} \frac{\rho}{\Mp^3} , \;\;\;
\ms^2 \simeq m_0^2=\frac{\beta^2 \Mp^2}{2 c_0} \gtrsim 100 \, {\rm mm}^{-2} , \;\;\;
\mbox{hence} \;\;\; c_0 \lesssim 10^{61} .
\label{vev-m-c1}
\end{equation}
Then, comparing with the value of $\lambda$ from Eq.(\ref{e:lambdadef}), we find
\begin{equation}
c_0 \lesssim \lambda^{-2} \;\;\; \mbox{hence} \;\;\; a_0 \lesssim 2 .
\label{c-lambda-EotWash}
\end{equation}

We recall in more details in appendix~\ref{sec:labo}
the various experimental constraints that can be derived
on the mass $\ms $ of the scalar field $\phi$, hence on the coefficient $c_0$.
We find that neutron interferometry cannot provide competitive constraints.
Casimir effect experiments give $m_\star^{-1} < 10^{3} \, {\rm mm}$ while
E\"ot-Wash experiments give the even more stringent upper bound
$m_\star^{-1} < 0.1 \, {\rm mm}$, leading to Eq.(\ref{c-lambda-EotWash}).

We also note in appendix~\ref{sec:astro} that cosmological
and astrophysical observations cannot constrain the fifth force associated with the scalar field
$\phi$ because of its very small range.
\section{Quantum stability}
\label{sec:Theoretical-constraints}


A priori the low energy cosmological constant $(\lambda \Mp)^4$ can
differ drastically from the one  in any ultraviolet
completion of the low energy theory under study.  Indeed contributions
from all the loop corrections of the low energy particles of the
standard model and the cosmological phase transitions would give large
effects to $\lambda$. In the following we take the curvature squared
$\mathcal R^2$ model to be the effective field theory at low energy
below all the standard model mass thresholds, and that the additional
contributions to the cosmological constant have been renormalised to
zero below this scale $\mu_0$. Hence at low energy the Universe is
only described by the effective action in~\eqref{R2-action}
where we assume that the other terms are either negligible or
irrelevant, i.e. suppressed by a high energy scale such as the string scale, although the higher
order terms are needed for the effective action to make sense and
avoid having ghost as we will discuss in section~\ref{sec:EFT}.
In cosmology, the dynamics is dictated by the one particle irreducible
action which is corrected by the one loop Coleman-Weinberg term
\begin{equation}
\delta V^{1-\textrm{loop}}(\phi)= \frac{\ms^4}{64\pi^2} \ln
\left(\frac{\ms^2}{\mu^2}\right)= \frac{\beta^4 \Mp^4}{256 \pi^2 c_0^2}\ln \left(\frac{\ms^2}{\mu^2}\right)
\label{1-loop-dV}
\end{equation}
at the renormalisation scale $\mu$ where $\ms$ is given in~\eqref{e:vev}. In cosmology as already noted the value of $\mu$ is not related to a collision energy like in particle physics experiment.
 It is not obvious which scale it corresponds to, hence its value can only be extracted  from observations.

Starting from a model with a vanishing cosmological constant corresponding to $V_{\rm vac}=0$  the
one-loop generated cosmological constant is
\begin{equation}
 (\lambda \,\Mp)^{4}= \frac{\beta^4 \Mp^4}{256 \pi^2 c_0^2}\ln \left(\frac{\ms^2}{\mu^2}\right) .
\end{equation}
which leads to  the relation
\begin{equation}\label{e:lower}
c_0\simeq  \frac{\beta^2}{16\pi }
\ln^{1/2}\left(\frac{\ms^2}{\mu^2}\right)\frac{1}{\lambda^2}\simeq
{10^{-2}\over \lambda^2}.
\end{equation}
Including the positive vacuum energy obtained for instance in a de Sitter string compactification $V_{\rm vac}>0$,  we conclude that~\eqref{e:lower} provides  a lower bound on $c_0$
\be
c_0\gtrsim \alpha \lambda^{-2}
\label{eq:c0-quantum}
\ee
where the coefficient $\alpha$ is an
undetermined model dependent coefficient.
Moreover the quantum corrections due to the scalaron cannot be arbitrarily low
since this would lead to a violation of the observational bound $c_i\lesssim \lambda^{-2}$. Hence the coefficient $c_0$ is both tightly bounded from above and below. This mechanism is reminiscent of what happens in the SLED proposal \cite{Aghababaie:2003wz}
where the fact that the extra dimensions have a typical size around 1~mm guarantees that
the quantum corrections due to the Kaluza-Klein modes do not exceed the present day value of the dark energy scale. It is also similar to the results obtained in~\cite{Upadhye:2012vh} for chameleon models.

In order to analyse the  quantum stability of the minimum of the
effective potential at $\phi_\star$  we also need to consider quadratic contributions in $\rho^2$  to
the  mass of the scalaron  from  leading
corrections for densities larger than the cosmological one
\begin{equation}
\ms^2 \simeq
\frac{\beta^2 \Mp^2}{2 c_0} - \frac{5 \beta^2\rho}{\Mp^2} + \frac{2 c_0 \beta^2 \rho^2}{\Mp^6}
+ .... \, ,
\end{equation}
where we neglected contributions arising from higher derivative
terms in the effective action~\eqref{R2-action}.
As explained in~\cite{Upadhye:2012vh}, the quantum stability of the minimum of the effective potential
is guaranteed by imposing the following conditions
\begin{equation}
\frac{1}{\rho} \frac{dm^6_\star}{d\rho}\le \frac{96 \pi^2 \beta^2}{\Mp^2}, \qquad \frac{d^2m^6_\star}{ (d\rho)^2}\le \frac{96 \pi^2 \beta^2}{\Mp^2}.
\end{equation}
This is always easily satisfied for large $c_0$ and densities $\rho \gg 10^{-4} \rho_{\Lambda 0}$.
Notice that the mass
$m_\star\propto \Mp/\sqrt{c_0}$ receives a negligible contribution from
the energy density $\rho$ which differs from the case
in~\cite{Upadhye:2012vh} where $m_\star\propto (\rho/\Mp)^{1/3}$.

Hence we conclude that the quantum correction due to the scalaron are under control provided that $c_0 \gtrsim \alpha \lambda^{-2}$, from Eq.(\ref{eq:c0-quantum}).
Together with the upper bound (\ref{c-lambda-EotWash}) this gives the finite range
\begin{equation}\label{c0-bounds}
\alpha \lambda^{-2} \lesssim c_0 \lesssim \lambda^{-2},
\end{equation}
where $\alpha \lesssim 1$ is not directly calculable.
This suggest that there is a relationship
\begin{equation}
c_0 \simeq \lambda^{-2}
\label{result}
\end{equation}
up to an undetermined numerical coefficient. In particular this implies that the mass of the scalar is of order
\be
\ms^2  \simeq \alpha^{-1} (\lambda\,\Mp)^2
\ee
corresponding to a range of order $\alpha^{1/2}\times 82$ microns.

Notice that the previous result (\ref{result}) hinges upon two assumptions. The first one is that the only contribution to the vacuum energy from quantum corrections originates
from the scalaron. If $c_0\sim c_2$, the massive graviton would also contribute in a similar way and would not change the order of magnitude of the loop correction. When the mass
$m_2$ of the spin 2 field is large, typically larger than $\mu_0$, its effect on the low energy vacuum is discarded as compensated by the cosmological constant counter term in (\ref{tune}).
We must also require that $V_{\rm vac}\ge 0$ otherwise the lower bound on $c_0$ does not hold anymore. This will direct us towards Minkowski or de Sitter compactifications in string theory.

So far we have decoupled the effects of matter fields on the vacuum energy by requiring that (\ref{tune}) holds. It is important to check that the mass of the scalaron does not suffer from another
hierarchy problem. In the Einstein frame, the masses of particles become field dependent
\be
m_i(\phi)= A(\phi) m_i
\ee
and all the energy scales such as $\mu$ are similarly multiplied by $A(\phi)= e^{\beta \phi/M_{\rm Pl}}$. For $\phi\ll M_{\rm Pl}$, the quantum corrections to the mass of the scalaron coming from  the quantum fluctuations of the matter particles arise from
\be
\delta V(\phi)= \sum_{j} (-1)^{2j} (2j+1)\theta(m_j-\mu_0) \frac{m_j(\phi)^4}{64\pi^2 }\ln \frac{m_j^2}{\mu^2}
\ee
where the ratio $m_j/\mu$ is $\phi$-independent. Expanding to second order in $\phi$ leads to the correction to the mass due to matter loops.
The vacuum energy also comprises the contribution from phase transitions and the bare cosmological constant
\be
\delta \tilde V (\phi)= A^4(\phi)\left(\tilde \Lambda_0 M_{\rm Pl}^2 +\rho_{\rm trans}\right).
\ee
The correction to the mass of the scalaron is obtained by expanding $\delta V(\phi) + \delta \tilde V(\phi)$ to second order. Now using (\ref{tune}) we have
\begin{align}
\delta V (\phi) + \delta \tilde V (\phi)&= A^4(\phi)\left (\tilde
  \Lambda_0 M_{\rm Pl}^2 +\rho_{\rm trans}+\sum_{j} (-1)^{2j}
  (2j+1)\theta(m_j-\mu_0) \frac{m_j^4}{64\pi^2 }\ln
  \frac{m_j^2}{\mu^2}\right )\\
&= 0 \nonumber
\end{align}
implying that the mass of the scalaron is not corrected by matter loops. Indeed the tuning of the vacuum energy in the matter sector is enough to cancel the quantum corrections to the mass of the scalaron due to matter loops.

\section{Effective field theory models}\label{sec:EFT}

The curvature squared models considered above are part of a low-energy
effective action which should be embedded in theories valid to higher
energy. Moreover they are not quantum mechanically consistent as a
consequence of the truncation of the effective action, i.e. there is a massive spin 2 ghost.  In this
section we consider different effective field theory models.  We
consider the curvature squared models as part of the low-energy
effective action of string theory compactification models
(see~\cite{Alvarez-Gaume:2015rwa} for a recent discussion of $R^2$
models in string theory), and supergravity extensions of the curvature
squared terms.


\subsection{String compactification}
\label{sec:string}

In this section we
consider compactifications of the effective action of string theory.
The ten dimensional perturbative effective action for type II strings reads in the
string frame~\cite{Green:1997tv,Green:1997di,Kiritsis:1997em,Antoniadis:1997eg,Peeters:2000qj}
\begin{multline}
  \label{e:EFTstring10d}
\mathcal S^{\rm string}_{10d}={1\over (2\pi)^7 \ell_s^8} \int\, d^{10}x \sqrt{-g}
 \Big({1\over g_s^2}{\cal R}_{(10)} + {2\zeta(3)\over 3\cdot
   2^7}{\ell_s^6\over g_s^2}
(t_8t_8-{1\over 8} \epsilon_{10}\epsilon_{10}) R^4 \cr
+  {4\zeta(2)\over 3\cdot     2^7} {\ell_s^6\over g_s^2}
(t_8t_8 \pm   {1\over 8} \epsilon_{10}\epsilon_{10}) R^4 +\cdots \Big),
\end{multline}
with a positive sign for type IIb string and a negative for type IIa string, and
where $\ell_{s}$ is the string length and $g_s$ the string coupling
constant. For maximally superstring theory the first quantum
correction to the Einstein-Hilbert action is the eight derivative
$R^4$ term. The dots stand for higher derivative terms and
non-perturbative contributions. We will be  working
in the weak coupling regime $g_s\ll1$ keeping only the tree-level contributions.
 The eight-index $t_8$ is defined
in~\cite[App.~B.2]{Peeters:2000qj}, and $\epsilon_{10}$ is the
totally antisymmetric rank ten tensor.

In this section we focus on non-warped compactifications in the absence of fluxes and of branes.
Compactifications of the type~II string on the background
$\mathbb R^{1,3}\times \mathcal M_6$ given by the product of four
dimensional flat space times a six dimensional manifold $\mathcal M_6$, i.e. with no warping,
leads to the parity-even effective action in four dimensions
\begin{equation}  \label{e:EFTstring4d}
\mathcal      S^{\rm string}_{4d}= \int d^{4}x \sqrt{-g} \left( -\lambda^4 \Mp^4+  {\Mp^2 \over2}\mathcal
        R + c_3 R_{\mu\nu}{}^{ba}R_{\nu\mu} {}^{ba}+c_4 \epsilon^{\mu\nu\rho\sigma}\epsilon_{abcd}
R_{\mu\nu}{}^{ab}R_{\rho\sigma}{}^{cd}+\cdots\right)
\end{equation}
 The curvature squared terms arising by integrating two powers the Riemann curvature  along the compactification manifold
lead to the parity-even $R^2$ terms in four
dimensions~\cite{Antoniadis:1997eg,Bachas:1999um,Antoniadis:2002tr}
\begin{equation}\label{e:cistring}
  c_3, c_4\sim  {1\over g_s^2\ell_{s}^2}
  \int_{\mathcal M_6}d^6x \sqrt{-g}  R^2\,.
\end{equation}%
where the $R^2$ here are either $R_{\mu\nu}{}^{ab}R_{\nu\mu}{}^{ba}$
or
$\epsilon^{\mu\nu\rho\sigma}\epsilon_{abcd}
R_{\mu\nu}{}^{ab}R_{\rho\sigma}{}^{cd}$ integrated along the internal
directions.
 Using the identity that $R_{\mu\nu}{}^{ab} R_{\nu\mu}{}^{ba}=
4(R_{\mu\nu})^2-\mathcal R^2+1/4\epsilon^{\mu\nu\rho\sigma}\epsilon_{abcd}
R^{ab}_{\mu\nu}R^{cd}_{\rho\sigma}$  giving Ricci squared term up to
the Gauss-Bonnet contribution  we can map the curvature
square terms in~\eqref{e:EFTstring4d} to the one in~\eqref{EH-R2}
using
\begin{align}
  &\int d^4x \sqrt{-g}\, \left( c_3 R_{\mu\nu}{}^{ba}R_{\nu\mu} {}^{ba} +
    c_4 \epsilon^{\mu\nu\rho\sigma}\epsilon_{abcd}
    R_{\mu\nu}{}^{ab}R_{\rho\sigma}{}^{cd}\right)\\
\nonumber &=  \int d^4x \sqrt{-g}\, \left( {c_3\over3} \mathcal R^2 +4c_3
  \left((R_{\mu\nu})^2-\frac13 \mathcal R^2\right)+
   \left(c_4+{c_3\over4}\right) \epsilon^{\mu\nu\rho\sigma}\epsilon_{abcd}
    R_{\mu\nu}{}^{ab}R_{\rho\sigma}{}^{cd}\right)\,.
\end{align}
One can always perform a field redefinition $g_{\mu\nu}\to g_{\mu\nu}+\alpha
R_{\mu\nu}+\beta \mathcal R\, g_{\mu\nu}$ which changes the coefficient of
the $R_{\mu\nu}^2-\frac13\,\mathcal R^2$ term as $4c_3\to 4c_3+\alpha$
and the coefficient of the $\mathcal R^2$ term as ${c_3\over3}\to
{c_3\over3}+ {\alpha+3\beta\over3}$~\cite{Tseytlin:1986zz,Forger:1996vj}.
Such a field redefinition of the metric is a
symmetry of the $S$-matrix and cannot change the spectrum of the
theory, therefore the signs of the curvature squared coefficients will
be compatible with the absence of tachyon as we have expanded around a
true vacuum of string theory which is therefore stable. Such field
redefinition must be compatible with the background configuration around which the perturbative action has been obtained. Although we do not elaborate on this here, one
can imagine that the  supersymmetry breaking terms considered in
section~\ref{sec:nilpotent}, where we only work at the $\mathcal N=1$ supergravity level,  when involving branes and anti-branes allow for a large  $\beta\gg\alpha$ in agreement with
the desired hierarchy~\eqref{e:hierarchy}.

The cosmological constant can arise by integrating the
$t_8t_8R^4$ over the internal manifold
\begin{equation}\label{e:lambdastring}
  V_{\rm vac}= {1\over g_s^2\ell_s^2}
  \int_{\mathcal M_6} d^6x  \sqrt{-g}
 t_8t_8R^4\,.
\end{equation}
For any compactification with a constant killing spinor the integral
$\int_{M_6} (t_8t_8-\epsilon_{10}\epsilon_{10}) R^4=0$
vanishes~\cite{Gross:1986iv}. Since $\epsilon_{10}\epsilon_{10} R^4=0$
in six dimensions this implies the vanishing of the integral of
$t_8t_8R^4$. To achieve a non vanishing cosmological constant $V_{\rm vac}\ne 0$
from~\eqref{e:lambdastring} one should have a compactification without constant killing
spinor by adding
brane and antibranes~\cite{Kachru:2003aw} or fluxes~\cite{Balasubramanian:2005zx} are both added to the setting leading to the
type of low energy models of the section~\ref{sec:sugra}, see also \cite{Sethi:2017phn}.

The four dimensional Planck mass is obtained  by integrating
three powers of the Riemann tensor along the compactification manifold
$\mathcal M_6$~\cite{Antoniadis:2002tr}
\begin{equation}\label{e:Mpstring}
  \Mp^2= {v_6\over (2\pi)^7\ell_{s}^2g_s^2}+ {1\over 3\cdot 2^7
    \,(2\pi)^7} {2\zeta(3)\over g_s^2\ell_{s}^2}
  \int_{\mathcal M_6}d^6x \sqrt{-g} \epsilon_6\epsilon_6 R^3
\end{equation}
with $v_6=V_6/\ell_s^6:=l_6^6/\ell_s^6$ the volume of $\mathcal M_6$ in unit of the string length. The second term is a correction to the first terms as long as $v_6 \gg 1$.

In a similar vein we can estimate the
effects of the higher derivative corrections to
the string theory effective action.
The tree-level higher derivative expansion to the string theory
effective actions is given by
\begin{equation}\label{e:exp4d}
  S_{10d}^{\rm string} = {1\over(2\pi)^7}\,\int d^{10}x \, \sqrt{-g} {1\over
    g_s^2\ell_s^8}\, \left(\mathcal R  + \alpha_3 \ell_s^6
    R^4+\alpha_5\ell_s^{10} R^6+\cdots\right)
\end{equation}
where the $\alpha_n$ are numerical coefficients. After
compactification this leads to

\begin{equation}\label{e:exp4dbis}
 S_{4d}^{\rm string} =\int d^4x \, \sqrt{-g}
  \biggl \lbrace V_{\rm vac} + \Mp^2 \biggl [  \frac{\mathcal R}{2}  + R^2 \ell_s^2 \, \sum_{p=0}^{\infty}
  d_p  ( \ell_s^2 R )^p  \biggl ] \biggl \rbrace   \, .
\end{equation}
In the case of non-warped compactification with no branes and anti branes, we have $V_{\rm vac}\equiv 0$ by supersymmetry.
The structure of the other terms  in the action follows an easy book-keeping principle.
The Planck mass is set by
\be
M_{\rm Pl}^2 \simeq \frac{l_6^6}{g_s^2\, \ell_s^8}\left(1+ \sum_{n\geq 3} \alpha_n
\frac{\ell_s^{2n}}{\tilde{l}_6^{\;2n}}\right) , \;\;\;
\mbox{with} \;\;\;
\int \frac{d^6 x}{V_6} \sqrt{-g} R^n  \sim \tilde{l}_6^{\;-2n} \, .
\ee
Here we defined $\tilde{l}_6^{\;-2}$ as the typical averaged value of $R$ over the six dimensional compactification, one typically expects $\ell_s \lesssim \tilde{l}_6 \lesssim l_6$.
The infinite tower of contributions is suppressed when ${\tilde l}_6 \gg \ell_s$,
which is thus a necessary condition for the expansion over $R$ of the four dimensional action.
The condition ${\tilde l}_6 \gg \ell_s$ is actually natural, as ${\tilde l}_6 \lesssim \ell_s$
would require extremely large curvature regions that would dominate the
integrals (this might be the case for orbifold compactifications or non-compact Calabi-Yau compactifications). On the other hand, this also suggests $l_6 \gg \ell_s$.
The higher order terms are
\be
d_p \simeq \sum_{n\geq \max(p+1,3)} \alpha_n \left(\frac{\ell_s}{\tilde{l}_6}\right)^{2(n-p-1)} \, ,
\ee
and the series are dominated by the first term in the same regime, ${\tilde l}_6 \gg \ell_s$.
This gives
\be
d_0 \simeq \alpha_3 \left(\frac{\ell_s}{\tilde{l}_6}\right)^4 , \;\;\;
d_1 \simeq \alpha_3 \left(\frac{\ell_s}{\tilde{l}_6}\right)^2 , \;\;\;\;
p \geq 2: \;\; d_p \simeq \alpha_{p+1} .
\ee
For the $R^2$ models to be a valid description of the physics
at low energy up to the string scale $M_s=1/\ell_s$, where the higher order terms appear
one must
require\footnote{The present construction  indicates that the relation between the ghost free
bi-metric massive gravity model~\cite{Hassan:2011zd} and
 the conformal curvature squared model
worked out in~\cite{Hassan:2013pca} could be valid beyond the
classical approximation. Although the small mass of the spin-2
Pauli-Fierz would seem  to  forbid the possibility of integrating
it out, the  construction of this section could give some indication
on constructing an ultraviolet completion for such massive gravity
theories involving string theory, see~\cite{deRham:2017xox} for some
recent discussions. [We would like to thank Andrew Tolley
for a discussion on this point.] }
that $d_0 \gtrsim d_p$ for all $p \geq 1$. This implies $\tilde{l}_6 \lesssim \ell_s$,
which contradicts the condition ${\tilde l}_6 \gg \ell_s$ for these expansions to be valid.

The $R^2$ model only applies up to the curvature scale $R \leq \tilde{l}_6^{-2}$, beyond wich
the terms $R^3$ and $R^4$ become greater in the action (\ref{e:exp4dbis}), and higher order terms
successively appear at higher curvature.
However, this perturbative cutoff $R \leq \tilde{l}_6^{-2}$ is typically much greater than practical
curvature scales. Notice that at this scale the linear term ${\mathcal R}$ still dominates
in the action (\ref{e:exp4dbis}).
These estimates give for the coefficients $c_i$ and the masses $m_i$
\be
c_i \sim \frac{1}{g_s^2} \left( \frac{l_6}{\ell_s} \right)^6 \left( \frac{\tilde{l}_6}{\ell_s}\right)^{-4} ,
\;\;\;
m_i^2 = \frac{M_{\rm Pl}^2}{c_i} \sim \ell_s^{-2} \left( \frac{\tilde{l}_6}{\ell_s} \right)^4 \, .
\label{ci-estimate}
\ee
We can see that the mass $m_2$ of the spin 2 ghost is above $\ell_s^{-1}$, which is beyond the
cutoff $\tilde{l}_6^{-1}$ of the $R^2$ truncation. Therefore, as could be expected,
the ghost is only due to the truncation and does not appear at scales where the
$R^2$ model applies.
From $\tilde{l}_6 \gg \ell_s$ we obtain $c_i \ll g_s^{-2} (l_6/\ell_s)^6$ hence
$c_i \ll M_{\rm Pl}^2 \ell_s^2$.
On the other hand, if we wanted to reach $c_0 \sim \lambda^{-2} \simeq M_{\rm Pl}/H_0$,
this would give $\ell_s^2 \gg 1/(M_{\rm Pl} H_0)$, hence $M_s \ll 10^{-3} {\rm eV}$.
This value for the string scale is clearly too low, as it is much below the scales probed
by colliders. Therefore, the coefficients $c_i$ obtained from the string effective action~\eqref{e:exp4d} must much below the value $\lambda^{-2}$ advocated in Eq.(\ref{result}).

As a result the $c_0$ term must have a different physical origin leading to  $c_0 \gg c_2$.
Provided that such a large value of $c_0$ can be achieved, this would keep the spin 2 ghost harmless, as explained above, while providing
a non-negligible scalaron term $c_0 {\mathcal R}^2$ and possibly the expected order of magnitude for the vacuum energy.
In the following we will consider what a warped compactification scenario could lead to, i.e.
we will analyse  the dynamics of the $T$ modulus corresponding to the volume of the compactification coupled to the supersymmetry breaking due to anti branes (See~\cite{Covi:2008ea} for
various constraints on such models). We will find that in these models the value of $c_0$ can be large compared to $c_2$.

\subsection{String-inspired supergravity  embedding}\label{sec:sugra}

The $R^2$ models can be trusted as low energy field theories provided there is a large hierarchy of scales between the scalaron mass scale $m_0$ and the string scale. Moreover
the ghost-like spin 2 graviton is troublesome and must be decoupled with $c_2 \ll c_0$. This can be achieved if the orign of the $c_0$ term is physically very different from
the way $c_2$ is generated. In particular, in the previous section we have seen that
$c_2 \ll M_{\rm Pl}^2 \ell_s^2 \ll \lambda^{-2}$ and $m_2 \gg \ell_s^{-1}$.
As we describe below, the coefficient $c_0$ can be increased up to $c_0 \sim \lambda^{-2}$,
while keeping $c_2$ fixed. This can be achieved by considering the dynamics of supersymmetry breaking involving the $T$ modulus. In this case, the scalaron is
 identified with the volume modulus of the string compactification and its low energy action is written in terms of a two derivative N=1 supergravity action, i.e. it does not involve
higher derivative terms and does not modify the value of $c_2$. In this section, we  describe the type of supergravity models inspired by warped compactifications which  leads to a large value of
$c_0$ compared to $c_2$.

 Here we first start by studying the dynamics
of the shape modulus $T$ determining the volume of the compactification manifold.
We then consider adding another  superfield $S$ which could represent a string
field associated to the breaking of supersymmetry by antibranes.

\subsubsection{Single superfield model}\label{sec:single}

In this section we consider a toy model involving
only a single dynamical modulus $T$.  We will show that this
model has negative vacuum energy and leads to the relation
$|c_0|\propto \lambda^{-4}$ which is in direct violation of the
current experimental bound on these coefficients as discussed in section~\ref{sec:observations}.

The model is determined by the K\"ahler potential
\begin{equation}
K=-3 \Mp^2 \ln \frac{T+\bar T}{\Mp} \, ,
\end{equation}
where the modulus $T$ is related to the compactification volume by
\be
\frac{T+\bar T}{\Mp}\simeq v_6^{2/3} \, .
\ee
The dynamics of the model is determined by the superpotential $W$. The superpotential should be determined from first principles but here we simply choose it for illustration of the type of physics one may expect. We consider the function depending on the parameters $M$ and $\alpha$
\begin{equation}
W= M^2\left(T - \alpha \Mp\right).
\end{equation}
For such a field the kinetic terms are given by
\begin{equation}
{\cal L}_{\textrm{kin}}= - K_{T\bar T} \partial T \partial \bar T ,
\end{equation}
while the scalar potential is
\begin{equation}
V(T,\bar T)= e^{K\over \Mp^2} \left(K^{T\bar T}\vert D_T W\vert^2 -3 \frac{\vert W\vert^2}{\Mp^2}\right) \, ,
\end{equation}
where we have defined
$D_T W= \partial_T W + K_T \frac{W}{\Mp^2}$.
This gives
\begin{equation}\label{e:Vttb}
V(T,\bar T)= \frac{2 M^4 \Mp^2}{3 (T+\bar T)^2} \,\left( 3\alpha - {T+\bar T\over \Mp} \right) \, .
\end{equation}
Redefining the real part of the scalar field as
\begin{equation}\label{e:TTbPhi}
T= \bar T= \frac{\Mp}{2} \, e^{-{\nu\phi\over\Mp}} \, ,
\end{equation}
the kinetic term and the potential read
\be
{\cal L}_{\textrm{kin}} = - \frac{3\nu^2}{4} (\partial\phi)^2 , \;\;\;
V(\phi) = \frac{2 M^4}{3}  e^{\nu\phi/\Mp}  \left( 3 \alpha e^{{\nu\phi\over\Mp}}-1\right ) \, .
\ee
The coupling to fermions in $\mathcal N=1$ supergravity is given by the field dependent mass
\begin{equation}
m_\psi\simeq  e^{K/2\Mp^2} \; m_\psi^{(0)} = e^{3\nu\phi\over 2\Mp} \, m_\psi^{(0)} \, ,
\end{equation}
where $m_\psi^{(0)}$ is the mass as appearing in the global supersymmetry case from the superpotential,
which identifies the coupling to matter.
The total scalar potential in the Einstein frame for  these supergravity models in the presence of matter is therefore
\begin{equation}\label{V-total-1}
V_{\textrm{total}}= e^{\nu\phi\over\Mp} \left[ \frac{2M^4 }{3}\left( 3 \alpha e^{\nu\phi\over\Mp} - 1\right) + \rho \, e^{\nu\phi\over 2\Mp} \right] \, .
\end{equation}
If we could discard the prefactor $e^{\nu\phi/\Mp}$, we would recover the potential
(\ref{e:Vscal}) associated with the $R^2$-model, in its scalar-tensor form
(\ref{S-phi}), by choosing $\nu=2\beta$ and $\alpha=2/3$. This would also recover
a canonically normalized kinetic term and give $c_0=-3\Mp^4/32 M^4 <0$,
$\lambda^4=-1/16 c_0$, with a negative value for $c_0$ and an exponent $a_0=4$
in Eq.(\ref{ci-lambda}) that is different from the preferred value $a_0=2$ of Eq.(\ref{a0-2}).
However, it is not possible to remove the prefactor by making a conformal transformation
$g_{\mu\nu} = e^{-\nu\phi/2\Mp} \tilde{g}_{\mu\nu}$.
Indeed, in this new frame the Planck mass becomes field dependent,
$\widetilde{M}_{\textrm{Pl}}=e^{-\nu\phi/4\Mp} \Mp$, and the additional contributions associated with the
transformation of the Ricci scalar change the scalar field kinetic term, which is no longer
canonically normalized.

Besides, at zero density the potential (\ref{V-total-1}) has a negative minimum,
$V_{\min}=-M^4/18\alpha$ if $\alpha>0$ (and is unbounded from below if $\alpha<0$).
In fact, at any finite density the minimum of the total potential is negative
(as $V_{\textrm{total}}\to 0^{-}$ for $\phi\to-\infty$).
Therefore, this model does not provide a positive vacuum energy and  is
not a realistic scenario.
In the next section, we will resolve
these shortcomings by constructing a de Sitter
$V_{\rm vac} >0$ model with a small vacuum energy  obtained  by introducing supersymmetry breaking fields.

\subsubsection{Nilpotent  supersymmetry breaking}\label{sec:nilpotent}

In this section we show how  to overcome the
 shortcomings of the previous model, i.e. a large and negative vacuum energy,  by introducing a second field $S$ satisfying the nilpotency constrained
\begin{equation}
(S-M_S)^2=0
\label{S-nilpotent}
\end{equation}
as befitting Goldstino multiplets which can arise for instance in the breaking of supersymmetry by anti D3-brane in string theory. We consider the scale $M_S$ as a parameter which will
 play a role in the vacuum energy. It is quite likely that by putting an anti-brane at the bottom of a warped throat of the warped compactification, the scale
 $M_S$ could be warped down exponentially like in a Randall-Sundrum scenario, i.e. $M_S \ll M_s$. We also introduce  ~\cite{Hasegawa:2015era} the K\"ahler potential which is extended to
\begin{equation}
K=-3 \Mp^2 \ln \left(\frac{T+\bar T}{\Mp}-\frac{S\bar S}{3\Mp^2}\right).
\end{equation}
Notice that the nilpotent field enters in the K\"ahler potential in a manner similar to  a matter field field $C$ on a D3 brane, i.e.
\begin{equation}
K=-3 \Mp^2 \ln \left(\frac{T+\bar T}{\Mp}-\frac{S\bar S}{3\Mp^2}-\frac{C\bar C}{3\Mp^2}\right).
\end{equation}
To leading order this leads  to kinetic terms in $\frac{|\partial C|^2}{T+ \bar T}$ for the associated matter scalar.
The superpotential is linear in the nilpotent field and assumed to be linear in the volume modulus. It would be extremely interesting to have a string construction of such a superpotential, here we choose for illustration
\begin{equation}
W=-W_0 +  M_WS  ( T -\langle T\rangle)\, .
\label{W-W0-S}
\end{equation}
where $W_0$ is real and positive such that $\frac{W_0}{M_{\rm Pl}}\gg
M_S M_W$. We have also chosen $\langle T\rangle \gtrsim M_{\rm Pl}$ to be another free parameter.
We focus on the case where $M_S \ll M_{\rm Pl}$ and $T\sim \langle T\rangle$.
The scalar potential is obtained in Eq.(\ref{V-general}) in Appendix \ref{sec:kah}. For
the superpotential (\ref{W-W0-S}) this gives
\be
V(T) = \frac{T+\bar T}{\Mp} \frac{M_W^2 M_S^2}{3\Delta^2} + \frac{2 M_S M_W W_0}{\Delta^2 \Mp} + \frac{M_W^2 | T - \langle T \rangle |^2}{\Delta^2}
- \frac{4 M_W^2 M_S^2 \Re\textrm{e}( T - \langle T \rangle))}{3 \Delta^2\Mp} ,
\label{V-T}
\ee
where the potential is evaluated at $S=M_S$, for the nilpotent case
(\ref{S-nilpotent}), and $\Delta$ is defined in Eq.(\ref{Delta-def}).
Because $T \sim \Mp$ and $W_0 \gg M_S M_W \Mp$, the second term is much greater than
the first term. The third term is much greater than the fourth term
if $| T - \langle T \rangle | \gg M_S^2/\Mp$.
Then, using $\Delta \simeq (T+\bar T)/\Mp$ as we assumed $M_S \ll \Mp$,
the scalar potential reads
\begin{equation}
V(T) \simeq \frac{2M_S M_W \Mp W_0}{(T+\bar T)^2}
+ \frac{M_W^2 \Mp^2 | T - \langle T \rangle |^2}{(T+\bar T)^2} .
\end{equation}
Putting
\be
T= \langle T\rangle\, e^{-{2\beta\phi\over\Mp}} e^{i\theta} \, ,
\ee
the potential is minimised for $\theta=0$ and $T$ is real and positive. The
field $\phi$ is canonically normalised, because $\beta=1/\sqrt{6}$,
and the scalar potential becomes
\begin{equation}
V(\phi) \simeq v_6 ^{-4/3}\frac{M_S M_W W_0}{2 M_{\rm Pl}} e^{4\beta\phi\over\Mp}
+ \frac{M_W^2\Mp^2}{4} \left(1 - e^{2\beta\phi\over\Mp} \right)^2 \, ,
\end{equation}
which corresponds to the $\mathcal R^2$ model (\ref{e:Vscal}) with
the following identifications
\begin{equation}
V_{\rm vac}= v_6^{-4/3}\frac{M_S M_W W_0}{2 M_{\rm Pl}} \, ;
\qquad c_0= \frac{\Mp^2}{4M_W^2}\,.
\end{equation}
The identification extends to the matter coupling as the normalised fermions $\psi_C$ associated to the matter field $C/\sqrt{T+\bar T}$ have a scalar dependent
mass term
\begin{equation}
{\cal L}_\psi\supset e^{K/2M_{\rm Pl}^2} (T+\bar T) \bar \psi _C \psi_C \propto
(T+\bar T)^{-1/2} \bar \psi _C \psi_C
\end{equation}
corresponding to a field dependent mass
\begin{equation}
m_\psi= e^{\beta \phi/{M_{\rm Pl}}} m_\psi^{(0)} \, .
\end{equation}
Notice that we have used the identification for the compactification
volume
\begin{equation}
\frac{\langle T\rangle}{M_{\rm Pl}} \simeq v_6^{2/3}.
\end{equation}
and we consider the regime $v_6 \gtrsim 1$, where the higher
order derivative interactions in the string effective action are
small.

We can now analyse the compatibility of this model with the bounds on
$c_0$ and $\lambda$.
First, the experimental upper bound on the range of the scalar field implies
$c_0 \lesssim \lambda^{-2}$ from Eq.(\ref{c-lambda-EotWash}).
Second, as $V_{\rm vac}\ge 0$, the lower bound~\eqref{eq:c0-quantum} from quantum stability
holds, $c_0 \gtrsim \lambda^{-2}$.
Thus, we obtain
\be
c_0 \sim \lambda^{-2}  \;\;\; \mbox{and} \;\;\; M_W \sim \lambda \Mp \simeq 0.1 \; {\rm meV} .
\label{c0-MW-1meV}
\ee
Moreover, adding to $V_{\rm vac}$ the loop correction coming from the
massive scalaron we have from Eq.(\ref{1-loop-dV})
\begin{equation}
\lambda^4 \Mp^4 = v_6^{-4/3}\frac{M_S M_W W_0}{2 M_{\rm Pl}}
+ \frac{\beta^4 M_W^4}{16\pi^2} \ln \frac{m_0^2}{\mu^2} \;\;\; \mbox{with} \;\;\;
m_0^2 = 2 \beta^2 M_W^2 \, ,
\end{equation}
which involves the infrared scale $\mu$ and we used Eq.(\ref{e:vev}) for $m_0$.
The one-loop contribution is already of order $\lambda^4\Mp^4$ and
$V_{\rm vac}$ must not be greater than $\lambda^4\Mp^4$. This gives
$\frac{W_0}{M_{\rm Pl}} \lesssim \frac{M_W^3}{M_S}$. Combining with our condition
$\frac{W_0}{M_{\rm Pl}}\gg M_S M_W$ this gives the range
\be
 M_S M_W \ll \frac{W_0}{\Mp} \lesssim \frac{M_W^3}{M_S} v_6^{4/3} ,
\end{equation}
and the measured vacuum energy can come from a combination of the de Sitter vacuum
energy of string theory and the one loop effect of the massive scalaron if the upper bound
on $W_0$ is reached.
In all cases, this requires $M_S\ll M_W v_6^{2/3}$, and for instance
$W_0\sim M_{\rm Pl} M_W^2$.
Finally, we need to check that the fourth term in Eq.(\ref{V-T}) is indeed negligible
as compared to the third term. This requires $\beta\phi/\Mp \gg M_S^2/\Mp^2 v_6^{2/3}$.
Using Eq.(\ref{e:vev}) this reads
$M_S^2 \ll (\rho/\rho_{\Lambda0}) v_6^{2/3} M_W^2$, which is always satisfied
when $M_S \ll M_W$.

  Finally  as $\phi\ll M_{\rm Pl}$ and $v_6 \gtrsim 1$
\begin{equation}
l_6 \gtrsim  \ell_s
\end{equation}
leading  to a massive spin 2 state with a mass
\begin{equation}
m_2 \gtrsim M_s \,,
\end{equation}
which decouples from the low energy dynamics. Hence at low energy the scalaron would be the only remnant of the compactification with a very low supersymmetry breaking scale corresponding to a vacuum energy of order ${\rm meV}^4$.

\section{Conclusion}
\label{sec:conclusion}

In this work we have considered that the low energy effective action describing the late time Universe is of pure gravitational origin and
requires no additional sector which would explicitly mimic dark energy. At low energy and low curvature, the most relevant terms in the action are quadratic in the curvature. We have  argued for the natural relation
$c_0\simeq \lambda^{-2}\simeq 10^{61}$ between the dimensionless
coefficients $c_0$ of the curvature squared terms and the
dimensionless (in Planck mass units) dark energy density.  We
have motivated this relation by  focussing on the experimental constraints which gives an upper bound $c_0\le\lambda^{-2}$, and the quantum stability of the model which requires
$c_0\gtrsim  \lambda^{-2}$. We have also shown how this relationship
can arise in various ultraviolet completions of the low energy model involving string compactifications with anti branes, guaranteeing that  higher curvature terms in the action can be safely neglected.

We have also  explained that the consistency of the low energy
effective action imposes the hierarchy $c_0\gg c_2$ between the
curvature squared terms. Only the precise equivalence between the Ricci
scalar squared term and a massive scalar field with a precise coupling to matter equal to $\beta=\frac{1}{\sqrt 6}$  allows one to increase
the value of the coefficient $c_0$ using  supersymmetry breaking terms.

The low energy ${\cal R}^2$ models can be mapped into scalar-tensor
theories with a scalaron of mass of the order $10^{-3}$~eV. Once embedded into plausible ultraviolet completions, this degree of freedom may be seen as what remains
at low energy from the physics of the ultraviolet completion. On the string
theory side, the scalar would arise as a compactification modulus. Irrespective of the ultraviolet completion, the low energy theory can be tested by laboratory experiments of the torsion pendulum type
where the scalar field would act as a fifth force. One can estimate the range of this new interaction to be of the order $1/\alpha^{1/2}\times 82$~microns where $\alpha$  should be of the order $\alpha \simeq 10^{-2}$. For such a range of order 10 microns,  Casimir experiments
with a drastically improved sensitivity could also be competitive. In conclusion, we expect that, if no additional degree of freedom is responsible for the late acceleration of the expansion, the detection of a new fifth
force should be around the corner. If such a fifth force failed to appear in the next generation of experiments, this would be a natural hint that some new physics, on top of gravity itself,  would be  responsible for the late time acceleration.

\section*{Acknowledgements}
\label{sec:acknowledgements}

We would like to thank John Donoghue, Ruben Minasian, Stefan Theisen and Andrew Tolley for useful discussions
and comments on this work. The research of
P. Vanhove has received funding the ANR grants QST ANR-12-BS05-003-01 and
``Amplitudes'' ANR-17-CE31-0001-01, and is partially supported by Laboratory of Mirror Symmetry NRU
HSE, RF Government grant, ag. N${}^\circ$ 14.641.31.0001.
P. Valageas is partly supported by the ANR grant ANR-12-BS05-0002.
This work is supported in part by the EU Horizon 2020 research and innovation programme under the Marie-Sklodowska grant No. 690575. This article is based upon work related to the COST Action CA15117 (CANTATA) supported by COST (European Cooperation in Science and Technology).
 The authors
are  grateful to the PSI$^2$ project of the Universit\'e Paris-Saclay
and the Institute d'Astrophysique Spatiale
(IAS) of the university of Orsay for its hospitality and its partial support during the completion of this work.
\appendix

\section{Laboratory experiments}
\label{sec:labo}

In these appendices we present some laboratory tests of modified
gravity by a fifth force~\cite{Jain:2013wgs}.

\subsection{Neutron Interferometry}
\label{sec:interferometry}

One of the tests of modified gravity which recently appeared is neutron interferometry~\cite{Brax:2013cfa}.
In brief, one compares the phase shifts due to the scalar field $\phi$
when a neutron takes two different paths and next interferes. The phase shift is given by
\begin{equation}
 \Phi= -\frac{m_N^2}{k} \int_{\rm path} \frac{\beta \phi}{M_{\rm Pl}} dz ,
 \label{shift-neutrons}
\end{equation}
where $m_N$ is the neutron mass, $k$ its momentum, and we integrate along
the trajectory. Typically, one considers the case when neutrons go through a planar cavity,
i.e. between two plates separated by a distance $2R$, or along the long axis $L$
of a cylindrical (or rectangular) cavity with diameter $2R$ with $2R \ll L$.
In general, this is compared with either the phase shift in the absence of a cavity, or with
the trajectory that grazes along the internal border of the cavity.
The latter is what we are going to investigate.
The equation of motion of the scalar field is
$-\partial^2\phi/\partial t^2 + \nabla^2\phi - \partial V/\partial\phi - \beta \rho/M_{\rm Pl}=0$,
and we consider static systems.
Both inside the plates and in the vacuum, where the density is not much
greater than $1 \, {\rm g.cm}^{-3}$, we have seen in Eq.(\ref{vev-m-c1})
that the field behaves like a coupled scalar with a mass $\ms^2$ that is
independent of the density, setting $\rho=0$ in~\eqref{e:vev},
\begin{equation}
\ms^2 = \frac{\beta^2 \Mp^2}{2 c_0} ,
\end{equation}
and a vacuum expectation value from~\eqref{e:vev}
\begin{equation}
\phi_\star(\rho) = - \frac{\beta \rho}{M_{\rm Pl} \ms^2}\, .
\end{equation}
Because $L \gg 2 R$, far from the entrance and exit sides of the cylinder the system
shows a cylindrical symmetry and the radial profile of the scalar field is set by the radius
$R$ of the cylinder. To obtain orders of magnitude, we can further simplify to a
one dimensional problem and consider the profile reached by the scalar field between two
plates at distance $2R$, which is given by
\begin{equation}
\frac{d^2 \phi}{dx^2}  - \ms^2 (\phi- \phi_\star(\rho) ) = 0 .
\end{equation}
Writing $\phi_{\infty}=\phi_\star(\rho_{\infty})$ and $\phi_0=\phi_\star(\rho_0)$,
where $\rho_{\infty} \sim 1 \, {\rm g.cm}^{-3}$ and $\rho_0 \ll \rho_{\infty}$
are the matter densities inside the plates and inside the vacuum chamber,
the even profile is given by
\begin{equation}
| x | > R : \;\;\; \phi= \phi_\infty + A e^{-\ms \vert x \vert} , \;\;\;
| x | < R : \;\;\; \phi= \phi_0 + B \cosh(\ms x) .
\label{phi-profile}
\end{equation}
Matching the solutions we find that
\begin{equation}
A = \frac{( \phi_0 - \phi_\infty ) e^{\ms R} \sinh(\ms R)}{\cosh(\ms R)+\sinh(\ms R)} , \;\;\;
B = - \frac{( \phi_0 - \phi_\infty )}{\cosh(\ms R)+\sinh(\ms R)} .
\end{equation}
Therefore, the difference between the field values along the central axis of the cavity and
its inner boundary is
\begin{equation}
\phi(0) - \phi(R) = ( \phi_0 - \phi_\infty ) \frac{\cosh(\ms R)-1}{\cosh(\ms R)+\sinh(\ms R)} .
\end{equation}
Using $\rho_\infty \gg \rho_0$, hence $|\phi_\infty| \gg |\phi_0|$,
the phase difference between the paths along the central axis and along the inner border
of cavity reads from Eq.(\ref{shift-neutrons})
\begin{equation}
\Delta \Phi = - \frac{\beta^2 m_N^2 \rho_\infty L}{ k \Mp^2 \ms ^2} \,
\frac{\cosh(\ms R)-1}{\cosh(\ms R)+\sinh(\ms R)} ,
\end{equation}
where $L$ is the length of the trajectory.

In practice we have $L \simeq 9$ cm and the wavelength is about 2.7
Angstroms~\cite{Lemmel:2015kwa}.
For masses of order $\ms \sim 10^{-11} \, {\rm GeV} = 10^2 \, {\rm cm}^{-1}$, this gives
$| \Delta \Phi | \lesssim 10^{-8}$, which is negligible.
Therefore, neutron interferometry is not competitive with E\"ot-Wash experiments
to constrain such scalar field models of such a mass.


\subsection{Casimir effect}
\label{sec:Casimir}

We now turn to the Casimir effect~\cite{Lamoreaux:1996wh}, associated with the scalar energy stored between
two plates of distance $2R$. As shown \cite{Brax:2014zta}, the scalar field, coupled to matter, yields an
additional contribution to the Casimir pressure felt by the plates, given by
\begin{equation}\label{cas}
\left| \frac{F_\phi}{A}\right| = V_{\rm eff}(\phi(0)) - V_{\rm eff}(\phi_0) ,
\end{equation}
where $A$ is the surface area of the plates and $V_{\rm eff}$ is again the effective potential,
$V_{\rm eff}(\phi)= \frac{1}{2} \ms^2 (\phi - \phi_\star)^2$.
This is the difference between the potential energy in vacuum (i.e., without the plates)
and in the vacuum chamber (i.e., with the plates at distance $2R$).
From the scalar field profile (\ref {phi-profile}) this is
\begin{equation}
\left\vert \frac{F_\phi}{A}\right\vert = \frac{1}{2} \ms^2 B^2 = \frac{\ms^2 (\phi_0 -\phi_\infty)^2}
{2 \,(\cosh(\ms R)+\sinh(\ms R))^2} .
\end{equation}
Using again $\rho_\infty \gg \rho_0$, this gives
\begin{equation}
  \left\vert \frac{F_\phi}{A}\right\vert = \frac{\beta^2
    \rho_\infty^2}{2 \Mp^2 \ms^2 \,( \cosh(\ms R)+\sinh(\ms R))^2}
\simeq \frac{\beta^2 \rho_\infty^2}{2 \Mp^2 \ms^2} \, e^{-2\ms R} ,
\end{equation}
where the last approximation, which holds both for small and large $mR$,
makes explicit the Yukawa suppression of the interaction between the two plates,
because of the finite range $m_\star^{-1}$ of the scalar field.

The most stringent experimental constraint on the intrinsic value of the Casimir pressure
has been obtained with a distance $2R=746$ nm between two parallel plates and reads
$\vert \frac{\Delta F_{\phi}}{A}\vert \le 0.35$ mPa~\footnote{ We have the conversion $1\rm { mPa}= 1.44 \ 10^6 (\lambda\,\Mp)^4$.} \cite{Decca:2007yb}.
The plate density is of the order of $\rho_\infty=10\ {\rm g.cm^{-3}}$.
This gives $\ms e^{\ms R} > 10^{-16} \, {\rm GeV}$, hence $\ms > 10^{-16} \, {\rm GeV}$,
which corresponds to a fifth-force range $\ms^{-1} < 10^3 \, {\rm mm}$.


\subsection{E\"ot-Wash experiment}
\label{sec:Eot-Wash}

The E\"ot-Wash experiment \cite{Adelberger:2003zx} involves two plates separated by a distance $D$ in which holes of radii $r_h$ have been drilled regularly on a circle. The two plates rotate with respect to each other. The gravitational and scalar interactions induce a torque on the plates which depends on the potential energy of the configuration. The potential energy is obtained by calculating the amount of work required to approach one plate from infinity \cite{Brax:2008hh,Upadhye:2012qu}. Defining by $A(\theta)$ the surface area of the two plates which face each other at any given time, a good approximation to the torque is obtained as the derivative of the potential energy of the configuration with respect to the rotation angle $\theta$ and  is  given by
\begin{equation}
T \sim a_\theta \int_D^{\infty} dx \; \frac{\Delta F_{\phi}}{A}(x)  \, ,
\end{equation}
where $a_\theta=\frac{dA}{d\theta}$ depends on the experiment.
Using the previous expression (\ref{cas}) for the Casimir energy we find that the torque is given by
\begin{equation}
T= a_\theta \
\frac{\beta^2\rho_\infty^2}{2 M_{\rm Pl}^2 \ms^3} e^{-\ms D} ,
\end{equation}
which is exponentially suppressed with the separation between the two plates $D$.
This leads to a constraint on the mass $m$ of the scalar field.

For the 2006 E\"ot-Wash experiment~\cite{Kapner:2006si}, we consider the bound obtained
for a separation between the plates of $D=55\mu{\rm m}$, (see also \cite{Perivolaropoulos:2016ucs,Antoniou:2017mhs} for a different analysis and new developments)
\begin{equation}
\vert T \vert \le a_\theta \Lambda_T^3 ,
\end{equation}
where $\Lambda_T= 0.35 \lambda\,\Mp$ \cite{Brax:2008hh}.
We must also modify the torque  in order to take into account the effects of a thin electrostatic shielding sheet of width $d_s=10\mu {\rm m}$ between the plates in the E\"ot-Wash experiment. This reduces the observed torque which becomes
$T_{obs}=e^{-m_c d_s} T$.
When the mass in dense media is very large, this imposes a strong reduction of the signal. For forces with a range of order $82$ microns, the reduction factor is small.
This E\"ot-Wash experiment gives a bound of
\begin{equation}
\ms\ge 1.22 \lambda\, \Mp,
\end{equation}
which is tighter than the one obtained from the Casimir experiment.

Hence we find that the present tests of the existence of fifth forces are within the right ballpark to test our models. In particular, if the distance between the plates could be increased, almost all
the natural range of values for $\ms$ with $\ms\simeq \lambda\,\Mp$  would be excluded.


\section{Astrophysical effects}
\label{sec:astro}

The curvature squared $\mathcal R^2$ models that we have considered
have a scalaron mass of order of $\lambda^2\Mp^2$ which implies that the range of the scalar interaction is tiny on cosmological scales. Hence
these models are ultra-local~\cite{Brax:2016vpd} whereby the only effects due to the propagation of the scalar happen for cosmological perturbations~\footnote{At the background level  we find that
${\cal R} \sim \frac{\vert T\vert }{M^2_{\rm Pl}}$ where $\vert T\vert \sim \rho $ is the trace of the energy-momentum tensor of matter with energy density $\rho$, as radiation does not contribute. The contribution in ${\cal R}^2$ is negligible compared to the Einstein-Hilbert term as long as $\vert T \vert \lesssim \lambda^2 M_{\rm Pl}^4$ corresponding to a scale factor $a \gtrsim \lambda^{2/3}\simeq 10^{-20}$ in the radiation era. Hence in the late radiation and matter eras, the ${\cal R}^2$ term plays no role at all and the background cosmology is equivalent to $\Lambda$-CDM. This is the regime that we consider here. }.

. Denoting by $\delta$ the density contrast of
cosmological perturbations, its growth for a wave number mode $k$ is given by
\begin{equation}
\ddot \delta + {\cal H} \dot H -\frac{3}{2} \Omega_m {\cal H}^2(1+ \epsilon)\delta=0
\end{equation}
where $\dot{}= d/d\eta$ and $\eta$ is the conformal time where the
Friedmann-Robertson-Walker metric is
\begin{equation}
ds^2= a^2(\eta)(-d\eta^2 +dx^2)
\end{equation}
and the conformal Hubble rate is
\begin{equation}
{\cal H}= \frac{\dot a}{a}.
\end{equation}
The matter fraction of the Universe is $\Omega_m$, presently around 0.25, and the modification of gravity by the scalaron leads to
\begin{equation}
\epsilon= 2\beta^2 \frac{k^2}{a^2 \ms^2}.
\end{equation}
For $\ms\sim \lambda\,\Mp$, and $k$ of cosmological interest, we have that $\epsilon\sim \frac{{\cal H}^2}{a^2\ms^2} \ll 1$ and no effect of the scalaron on large scale structure  can be inferred.

Similarly, the ultra-local models are such that the scalar field is at its minimum and the coupling becomes matter density dependent
\begin{equation}
A(\rho)=1  -2 c_0  \frac{\rho}{\Mp^4}\simeq 1  -2\lambda^2
\frac{\rho}{(\lambda\Mp)^4}
\end{equation}
where we used the relation $c_0\simeq\lambda^{-2}$.
The induced fifth force of astrophysical scale $F=-\nabla A(\rho)$ is negligible as $c_0 \lambda^4 \sim \lambda^2 \ll 1$ and the factor $\rho/(\lambda\Mp)^4$ cannot compensate being at most a few thousands in the core of galaxies.

\section{Scalar potential in $\mathcal N=1$ supergravity in four dimensions}
\label{sec:kah}

In this appendix we gather useful details about the scalar potential
for nilpotent models in $\mathcal N=1$ supergravity in four dimensions.
The scalar potential can be obtained using the master formula
\be
V= e^{K/M^2_{\rm Pl}} \left ( K^{I\bar J} D_I W \bar D_{\bar J} \bar W - 3 \frac{\vert W\vert^2}{\Mp^2}\right)
\ee
where $I,J $ label the superfields $\Phi^I$. $K$ is the K\"ahler potential which is a real function of the superfields and $W$ is the superpotential, an analytic function of the superfields
considered as complex variables. The covariant derivative is
\be
D_I W= \partial_I W + \frac{K_I}{\Mp^2} W
\ee
and subscripts denote derivations, e.g. $K_I=\frac{\partial K}{\partial \Phi^I}$. Here $K_{{I\bar J}}$ is the Hessian matrix of the K\"ahler potential and its inverse is $K^{\bar I J}$ such that $K^{\bar I J}K_{J\bar L}= \delta ^{\bar I}_{\bar L}$.

The nilpotent models are defined by the K\"ahler potential
\be
K=-3 \Mp^2 \ln \Delta
\ee
where we have defined
\be
\Delta= \frac{T+\bar T}{\Mp}-\frac{S\bar S}{3\Mp^2}.
\label{Delta-def}
\ee
We find
\be
K_T=-\frac{3\Mp}{\Delta},\qquad K_S= \frac{\bar S}{\Delta}
\ee
and the matrix
\be
K_{I\bar J}:= \frac{1}{\Delta^2} \left    ( \begin{array} {cc}
3& -\frac{ S}{\Mp}\\
-\frac{\bar S}{\Mp}& \frac{T+\bar T}{\Mp}\\
\end{array}
\right )
\ee
with its inverse
\be
K^{\bar I J} := \frac{\Delta }{3} \left    ( \begin{array} {cc}
\frac{T+\bar T}{\Mp}& \frac{S}{\Mp}\\
\frac{\bar S}{\Mp}& 3 \\
\end{array}
\right ).
\ee
As a result we have
\be
D_S W= \partial_S W + \frac{\bar S}{\Delta \Mp^2} \,W, \qquad  D_T W= \partial_T W -\frac{3}{\Delta \Mp} W
\ee
and
\begin{multline}
K^{I\bar J} D_I W \bar D_{\bar J} \bar W = \frac{(T+\bar T)}{\Mp}
\frac{\Delta}{3} \vert \partial_T W\vert^2 + \Delta \vert \partial_S
W\vert^2 +{2\Delta \over 3\Mp}\Re\textrm{e} (\partial_T
W\, \bar S\partial_{\bar S}\bar W)\cr
-2\frac{\Delta}{\Mp}\Re\textrm{e}(\bar W \partial_T W) +3\frac{\vert W\vert^2}{\Mp^2}.
\end{multline}
Notice that the ``no-scale" form of the K\"ahler potential associated to the non-compact space $SU(1,2)/U(1)^2$ implies that the terms in $\vert W\vert^2$ in the scalar potential cancel. This implies
that the scalar potential reads for a general superpotential
\be
V= \frac{(T+\bar T)}{\Mp} \frac{\vert \partial_T W\vert^2}{3\Delta^2} + \frac{ \vert \partial_S W\vert^2}{\Delta^2} +{2 \over 3\Mp\Delta^2}\Re\textrm{e} (\partial_T
W\, (\bar S\partial_{\bar S}\bar W-3\bar W)) \, .
\label{V-general}
\ee


\end{document}